\documentclass[prd,twocolumn,%
  showpacs,amsmath,amssymb]{revtex4-1}
\usepackage{graphicx}
\usepackage[breaklinks, colorlinks=true,linkcolor=blue, citecolor=cyan]{hyperref}
\usepackage{color}

\newcommand{\bv}{\boldsymbol{v}}
\newcommand{\bF}{\boldsymbol{F}}
\newcommand{\bp}{\boldsymbol{p}}
\newcommand{\bx}{\boldsymbol{x}}
\newcommand{\energy}{\varepsilon}
\newcommand{\benergy}{\bar{\varepsilon}}
\newcommand{\BEP}[1]{\bar{\varepsilon}_{#1}'}
\newcommand{\lnE}{x}
\newcommand{\dos}{\delta}

\begin{document}

\title{
Fixed points and flow analysis on off-equilibrium dynamics
in the boson Boltzmann equation}
\author{Kenji Fukushima}
\author{Koichi Murase}
\author{Shi Pu}
\affiliation{Department of Physics, The University of Tokyo, %
  7-3-1 Hongo, Bunkyo-ku, Tokyo 113-0033, Japan}

\begin{abstract}
  We consider fixed points of steady solutions and flow directions
  using the boson Boltzmann equation that is a one-dimensionally
  reduced kinetic equation after the angular integration.  With an
  elastic collision integral of the two-to-two scattering process, in
  the dense (dilute) regime where the distribution function is large
  (small), the boson Boltzmann equation has approximate fixed points
  with a power-law spectrum in addition to the thermal distribution
  function.  We argue that the power-law fixed point can be exact in
  special cases.  We elaborate a graphical presentation to display
  evolving flow directions similarly to the renormalization group
  flow, which explicitly exhibits how fixed points are connected and
  parameter space is separated by critical lines.  We discuss that
  such a flow diagram contains useful information on thermalization
  processes out of equilibrium.
\end{abstract}
\maketitle

\section{Introduction}

Understanding thermalization dynamics in quantum systems is a
long-standing and yet unresolved problem.  Even with modern computer
advances, solving the first-principle quantum field theories
numerically in Minkowskian spacetime demands not only enormous
computing resource but also algorithmic innovations.  We thus need to
assume a reduction of full quantum dynamics in some particular regimes
according to our interested problems.  In the dilute regime the most
useful and widely adopted approach is the Boltzmann equation that can
be in principle derived as a quasi-particle approximation of the full
quantum equation of motion, i.e.\ the Kadanoff-Baym or 2PI
equation~\cite{KB,Cornwall:1974vz} (see
Refs.~\cite{Blaizot:2001nr,Berges:2004yj} for comprehensive reviews).

In the context of the relativistic heavy-ion collision (see
Ref.~\cite{Fukushima:2016xgg} for a recent review on early
thermalization problems), with help from the Boltzmann equation for
gluon interactions, the thermalization time scale has been estimated
parametrically in terms of the strong coupling constant and the
momentum scale that characterizes the initial condition.  In this way,
the bottom-up thermalization scenario has been
established~\cite{Baier:2000sb}, which is further refined later in
Refs.~\cite{Kurkela:2011ti,Blaizot:2011xf}, and is still continued to
recent works~\cite{Kurkela:2015qoa}.  In fact, since the very early
days of the heavy-ion collision physics, the Boltzmann equations has
been the common theoretical tool for the investigation of
isotropization and thermalization, as pioneered in
Ref.~\cite{Baym:1984np} in the relaxation-time approximation and in
Ref.~\cite{Mueller:1999pi} with gluon-gluon scattering.  We note that
a conjecture on a transient formation of the Bose-Einstein
condensate~\cite{Blaizot:2011xf} inspired many numerical simulations
under an overpopulated condition~\cite{Berges:2012us,Xu:2014ega},
which may be significantly affected by full interaction
dynamics~\cite{Huang:2013lia,Berges:2016nru,Blaizot:2016iir}.

In the dense regime, the collision integral in the Boltzmann equation
involves higher order scattering processes, and it would make more
sense to solve the time evolution in terms of not particles but
fields.  Recent years, we have witnessed significant developments in a
method called the classical statistical simulation (CSS).  The CSS is
a semi-classical approximation to deal with quantum time evolution.
In fact, the Yang-Mills theory, that governs the fundamental laws of
gluon interactions, has rich (chaotic) contents even on the classical
level as discussed in Ref.~\cite{Biro:1993qc}.  Later, in
Ref.~\cite{Arnold:2005ef}, by solving the classical Yang-Mills theory
coupled with Vlasov equation (i.e.\ electromagnetic coupled Boltzmann
equation), the numerical results imply that a possible turbulent-like
energy cascade may help isotropization in weakly coupled non-Abelian
plasmas, which is a numerical demonstration of the Chromo-Weibel
instability scenario~\cite{Mrowczynski:1988dz,Arnold:2003rq} (see also
Refs.~\cite{Rebhan:2004ur,Rebhan:2005re} for semi-analytical
treatments of the non-Abelian plasma instabilities).  The energy decay
with similar power-law behavior has been discussed in
Ref.~\cite{Ipp:2010uy}, and see also Ref.~\cite{Fukushima:2013dma} for
transverse structure formation as well as the longitudinal power-law.
Alternatively, in Ref.~\cite{Mueller:2006up}, a different type of
power-law in the energy decay has been suggested for non-Abelian
plasmas.  Now, we should note that a longitudinally expanding case has
been studied intensively~\cite{Attems:2012js,Berges:2013eia}, which
generally tends to hinder isotropization.

Interestingly, in high-energy reactions, as a result of small-x
evolution of the parton distribution functions, the classical
treatment would be a good approximation, and the theoretical framework
is well founded under the name of the Color Glass Condensate (CGC)
(see Refs.~\cite{Gelis:2010nm,Blaizot:2016qgz} for reviews).  Then, an
instability has been discovered once the CGC coherent fields are
disturbed by quantum fluctuations~\cite{Romatschke:2005pm} (see also
Ref.~\cite{Berges:2004ce} for a related attempt to explain early
thermalization in the heavy-ion collision), which motivated systematic
investigations on the real-time Yang-Mills dynamics and led to a clear
recognition of non-Abelian wave
turbulence~\cite{Berges:2008mr,Berges:2013eia}, where the
``turbulence'' refers to a power-law spectrum with a certain value of
exponent~\cite{Micha:2004bv}.  Actually, the theoretical description
of the CSS is equivalent to what is called the non-linear
Schr\"{o}dinger equation, which is frequently used in the context of
the wave turbulence~\cite{nazarenko2011wave}.  Theoretically speaking
the exponent of the power-law spectrum may not be unique but take
different values depending on microscopic processes.  The typical
values of exponents correspond to the particle cascade and the energy
cascade.  It is also possible to anticipate an even larger
exponent~\cite{Carrington:2010sz}, and this concept of non-thermal
steady states is generalized as the non-thermal ``fixed point'' in
analogy to the Wilsonian renormalization-group (RG) flow.  For a
review on the non-thermal fixed point and scaling solutions, see
Ref.~\cite{Nowak:2013juc}.  In a similar sense to the RG analysis, the
universality has been pursued
numerically~\cite{Berges:2013fga,Berges:2014bba}, and also the scaling
law relations among critical exponents have been
investigated~\cite{Berges:2015ixa}.  All these recent progresses are
very nicely summarized in a lecture note~\cite{Berges:2015kfa}.

We note that the CSS has been highly elaborated and the Wigner
distribution function that encodes the initial fluctuations has been
determined for expanding
geometries~\cite{Fukushima:2006ax,Dusling:2011rz}.  The precise form
of the Wigner distribution function is crucial to reproduce the
perturbative results correctly, as argued to recover the Schwinger
mechanism formula~\cite{Gelis:2013oca}.  Then, one would be naturally
led to an idea that the CSS with the correct Wigner distribution may
already capture the Boltzmann dynamics and may achieve isotropization
and thermalization~\cite{Gelis:2013rba}.  In fact, there are
theoretical discussions on the relation between the Boltzmann equation
and the classical field equation~\cite{Mueller:2002gd,Jeon:2004dh}.
As long as the distribution function $f$ is large enough to satisfy
$f^3 \gg f^2$, these two equations could describe the same physics
equivalently.  Therefore, in this way, the Boltzmann study with
$f^3 \gg f^2$ can give an intuitive explanation for some features in
the CSS results~\cite{Epelbaum:2014mfa}.  One might think that
 $f^3 \gg f^2$ immediately implies that higher order scattering
processes should take part in the collision integral.  As we will
discuss later, however, there exists a certain coupling window in
which $f^3 \gg f^2$ but the lowest scattering process is still
dominant in the collision integral.

If we consider the simplest elastic $2\leftrightarrow 2$ scattering
and drop $f^2$ assuming $f^3 \gg f^2$, the Boltzmann equation with
such a truncation has multiple power-law fixed points, one of which
corresponds to the Rayleigh-Jeans approximated thermal distribution.
If we in turn drop $f^3$ in a dilute regime of $f^2 \gg f^3$, the
Boltzmann equation again accommodates several power-law fixed points.
The genuine quantum Bose-Einstein distribution function is the
asymptotic solution only with a combination of both $f^2$ and $f^3$.
The goal of this paper is twofold.  The first is to make a complete
classification list of the fixed points or the steady solutions in
such truncated Boltzmann equations in the dense and the dilute regimes
(see Ref.~\cite{Mehtar-Tani:2016bay} for similar analysis).
The second is to clarify the relevance of these fixed points in the
full quantum case with $f^2$ and $f^3$ (see
Ref.~\cite{Epelbaum:2015vxa} for a closely related work with a similar
motivation).  For the latter purpose we will propose a new graphical
representation of our results in such a way that resembles the RG
flows with fixed points and critical lines.  The great advantages of
such a representation include;  (1) intuitively understandable in
analogy to the RG flow, (2) clear to judge whether the fixed points
are attractive or repulsive, and (3) providing information on critical
lines.  In particular, we would emphasize that the recognition of the
critical lines is regarded as a novelty in our present work, though it
is a natural anticipation from the RG analogue.

This paper is organized as follows:  In Sec.~\ref{sec:boson} we make
an overview of the Boltzmann equation, especially one-dimensionally
reduced one after the angular integration.  Such a simplified
representation of the Boltzmann equation is specifically referred to
as the boson Boltzmann equation in the literature.  We then proceed to
the classification of the approximate fixed points or steady solutions
in Sec.~\ref{sec:fixedpoints}.  We will there find not only the
power-law solutions belonging to the Kolmogorov-Zakharov (KZ) scaling,
but also a new self-similar solution.  In Sec.~\ref{sec:flow} we
present our central results in a form of the flow diagram.  We make
clear the structure of distinct fixed points and critical lines.
Section~\ref{sec:conclusion} is devoted to the conclusion.  We note
that we set $\hbar=c=k_{B}=1$ for notational brevity.

\section{Boson Boltzmann equation}
\label{sec:boson}

We start with the ordinary Boltzmann equation,
\begin{equation}
  \frac{\partial}{\partial t}f_1
  + \bv\cdot\frac{\partial f_1}{\partial\bx}
  + \bF\cdot\frac{\partial f_1}{\partial\bp}
  = C[f]\;,
  \label{eq:BE_01}
\end{equation}
where $f_1=f_1(t,\bx,\bp_1)$ is the distribution function for particle
$1$ with its momentum $\bp_1$ or energy $\energy_1(\bp_1)$.  The
particle velocity and the external force are denoted by $\bv$ and
$\bF$, respectively, and $C[f]$ represents the particle scattering
effects, which is called the collision integral.  For weakly
interacting systems we can take account of the collision integral
perturbatively, and the elastic scattering at the lowest non-trivial
order is the two particle process, namely, the $2\leftrightarrow 2$
scattering.  At this order the collision integral generally reads:
\begin{equation}
  \begin{split}
  & C[f] = \frac{1}{2\energy_1} \int \prod_{i=2}^{4}
  \frac{d^3 p_i}{(2\pi)^3 (2\energy_i)} W(\{p_i,\energy_i\}) \\
  & \quad\times \bigl[(1+f_1)(1+f_2)f_3 f_4
  - f_1 f_2(1+f_3)(1+f_4)\bigr]
  \end{split}
\label{eq:collision_01}
\end{equation}
with particles $2$, $3$, and $4$.  In the above expression
$W\,d^3 p_3 d^3 p_4/(2\pi)^6$ represents the probability for the
scattering process from initial state particles $1,2$ to final state
particles $3,4$ within the phase space $d^3 p_3 d^3 p_4/(2\pi)^6$.  We
can express the probability using the scattering amplitude as
\begin{equation}
  \begin{split}
    W(\{p_i,\energy_i\}) &= |\mathcal{M}_{12\to 34}|^2\,
    \delta(\energy_1 + \energy_2 - \energy_3 - \energy_4) \\
    & \qquad \times\delta^{(3)}(\bp_1 + \bp_2 - \bp_3 - \bp_4)\;.
  \end{split}
\end{equation}
In this work we assume spatial homogeneity for the distribution
function to drop $\bx$ dependence hereafter.  We also consider a
special case with spherically symmetric momentum dependence, i.e.\ the
interaction has no angular preference.  This is the case for the
quartic vertex in the $\phi^4$ scalar theory, for example.  Thanks to
the symmetry, the kinetic equation simplifies to be one-dimensional,
which not only reduces the computational costs but also resolves the
subtle ambiguity on the energy-momentum conservation for discretized
momenta.

The symmetry requires that $\energy=\energy(|\bp|)$ and
$|\mathcal{M}_{12\to 34}|$ would be in general a function of momentum
modulus.  We will, however, introduce the density of states later, and
without loss of generality, we can take it as just a constant, i.e.,
$|\mathcal{M}_{12\to 34}|=g$.  Thus, we can carry out the angular
integration in $C[f]$ and the Boltzmann equation \eqref{eq:BE_01}
takes a simple form, which is often referred to as the boson Boltzmann
equation (for this Ref.~\cite{Markowich2005} contains a short review),
that is written as
\begin{equation}
  \rho_1 \frac{\partial f_1 (t,\energy_1)}{\partial t}
  = Q[f](\energy_1)\;,
\end{equation}
where we defined the density of states per space
volume~\cite{PhysRevA.56.575} from
\begin{equation}
  \rho_1 \equiv \int\frac{d^3 p}{(2\pi)^3}\,
  \delta\bigl( \energy_1-\energy(|\bp|) \bigr)\;.
\end{equation}
The simplified collision integral $Q[f]$ is a function of $\energy_1$
as follows:
\begin{equation}
  \begin{split}
    & Q[f](\energy_1) = \int d\energy_2\,d\energy_3\,d\energy_4\,
    S(\{\energy_i\}) \\
    & \;\times \bigl[(1+f_1)(1+f_2)f_3 f_4 - f_1 f_2 (1+f_3)(1+f_4)\bigr]\;,
  \end{split}
  \label{eq:Q}
\end{equation}
where in the latter part with $f_i$ the first term represents a
process of particles $3$ and $4$ scattering into particles $1$ and
$2$, and the second term is the inverse process from $1$ and $2$ into
$3$ and $4$.  The interaction kernel in our modeling convention is
parametrized as
\begin{equation}
  S(\{\energy_i\}) = g^2\,
  \delta(\energy_1 + \energy_2 - \energy_3 - \energy_4)\,
  \energy_{\min}^\alpha
  \label{eq:S_min_01}
\end{equation}
with $\energy_{\min}\equiv \min(\{\energy_i\})$.  The above form is
typical in theoretical models such as the $\phi^4$ theory (see
Refs.~\cite{Semikoz:1995rd,Berges:2015kfa} for examples).  We note
that $\alpha$ is a constant associated with the density of states (and
the momentum dependence in the amplitude).  For the relativistic case,
$\alpha=1$~\cite{Mehtar-Tani:2016bay}, while for the non-relativistic
case $\energy_i$ in the denominator of the phase  space volume is
replaced with the mass $m$, which leads to
$\alpha=1/2$~\cite{PhysRevA.53.381,Semikoz:1995rd}.  We give more
detailed discussions on $Q[f]$ and $S(\{\energy_i\})$ in
Appendix~\ref{sec:appA}.

\section{Fixed points}
\label{sec:fixedpoints}

The thermal distribution function should be the final destination of
the time evolution in the boson Boltzmann equation.  This is the
literal definition of ``thermalization'' and our central interest is
to clarify possible paths toward thermalization which would be
substantially affected by the structures of other fixed points and
flow patterns connecting or disconnecting them.

In numerical simulations power-law spectra have been found as
transient states on the way toward thermalization, and thus, for
theoretical characterization of thermalization, it would be the most
essential first step to understand those power-law spectra as much
analytically as possible.  For such analytical treatments the boson
Boltzmann equation provides us with a useful framework.  It is much
simpler than the original Boltzmann equation, and nevertheless, it
still encompasses a variety of steady solutions.  We shall first
summarize these analytical solutions in what follows below.

In this paper we limit ourselves to the $2\leftrightarrow 2$
scattering in the collision integral, and then discuss two
approximated forms in extreme regimes as well as the full quantum one
in Eq.~\eqref{eq:Q}.

The first extreme is the dense regime or we will call it the
$f^3$-regime in this paper.  In a kinetic region where $f$ is larger
than the unity, $f^3$ terms become dominant over $f^2$ terms in the
collision integral~\eqref{eq:Q}.  One may think that more and more
particles would be involved in the collision integral for $f\gg 1$,
but there is a certain kinetic window in which $f^3\gg f^2$ is
compatible with the truncation up to the $2\leftrightarrow 2$ process.
This is the case for
\begin{equation}
  1 \ll f(\energy) \ll g^{-2}\;,
\end{equation}
that holds at sufficiently weak coupling.  Then, in the $f^3$-regime,
the kinetic equation is approximated as
\begin{equation}
  \begin{split}
    \rho_1 \frac{\partial f_1}{\partial t}
      &= g^2 \int_{\energy_2,\energy_3,\energy_4} \!\!\!
        \delta(\energy_1+\energy_2-\energy_3-\energy_4)\,
        \energy_{\min}^\alpha \\
      &\quad\times
        \big( f_1 f_3 f_4 + f_2 f_3 f_4 - f_1 f_2 f_3 - f_1 f_2 f_4 \bigr)\;,
  \end{split}
  \label{eq:dense}
\end{equation}
where $\int_{\energy_{i}}\equiv \int d\energy_{i}$.

Another extreme is the dilute regime or we will call it the
$f^2$-regime.  If the momentum or energy is sufficiently large, the
distribution function should generally get smaller and eventually we
come to the kinematic regime where
\begin{equation}
  f^3(\energy) \ll f^2(\energy) \ll 1\;.
\end{equation}
Then, the kinetic equation in the $f^2$-regime is approximated as
\begin{equation}
  \rho_1 \frac{\partial f_1}{\partial t}
  = g^2 \int_{\energy_2,\energy_3,\energy_4} \!\!\!
  \delta(\energy_1 \!+\! \energy_2 \!-\! \energy_3 \!-\! \energy_4)\,
  \energy_{\min}^\alpha
    \big( f_3 f_4 - f_1 f_2\bigr)\;.
  \label{eq:dilute}
\end{equation}
In the subsequent subsections we will consider the steady solutions
for Eqs.~\eqref{eq:dense} and \eqref{eq:dilute}.

\subsection{Thermal distribution}
\label{sec:thermal}

It is well understood that the detailed balance is satisfied for the
thermal distribution function.  That is, we can readily confirm that
the thermal distribution function makes $Q[f]$ vanishing.  For the
full quantum case with both $f^2$ and $f^3$ terms, we can find the
solution from the famous $H$ theorem as
\begin{equation}
  f_{\rm T}(\energy) = \frac{1}{e^{\beta(\energy-\mu)}-1}\;,
  \label{eq:fT}
\end{equation}
where two parameters in the above Bose-Einstein distribution,
$\beta=1/T$ and $\mu$, represent the temperature inverse and the
chemical potential, which dynamically depends on the choice of
$f(\energy)$ at $t=0$.

In the $f^3$-regime, we can use the $H$ theorem in the same
way~\cite{Epelbaum:2014mfa} to find a thermal fixed point,
\begin{equation}
  f_{\rm RJ}(\energy) = \frac{1}{\beta(\energy-\mu)}\;.
  \label{eq:RJ}
\end{equation}
This is nothing but a Rayleigh-Jeans approximated form of the Planck
spectrum~\eqref{eq:fT} for $\energy-\mu\ll T$.  Indeed, in this region
for $\energy-\mu\ll T$, the Bose-Einstein distribution is infrared
enhanced, so that $f_{\rm T}(\energy)\approx f_{\rm RJ}(\energy)\gg 1$
as they should be in the $f^3$-regime.

In the $f^2$-regime, on the other hand, it is again immediate to find
a thermal fixed point as given by
\begin{equation}
  f_{\rm MB}(\energy) = e^{-\beta(\energy-\mu)}\;.
\end{equation}
This is the Maxwell-Boltzmann distribution in classical physics, that
is again a natural consequence from the fact that quantum effects are
negligible for dilute systems.  Here, let us make a remark on the
usage of the word, ``classical'', which is sometimes confusing in the
literature.  Usually $f_{\rm RJ}(\energy)$ is often referred to as
classical in a sense that this is a solution in the $f^3$-regime where
the semi-classical approximation works.  In fact, the CSS using the
classical equation of motion would lead to this power-law form of the
thermal spectrum.  Also, $f_{\rm MB}(\energy)$ is of course a classical
distribution in a conventional sense of statistical mechanics.

We point out that there is always another solution that satisfies the
detailed balance, that is the constant solution given by
\begin{equation}
  f(\energy) = \text{const}\;.
\end{equation}
This makes $Q[f]$ vanishing in all the $f^2$-regime, the $f^3$-regime,
and the full quantum regime.  The physical meaning is obvious;  this
is a trivial solution in the $\beta\to 0$ limit, and the integrated
total energy and particle number are ill-defined.  Thus, such a
constant solution has no physical significance.  In later discussions
on the flow diagram, however, we should be aware of the existence of
this solution in order to understand the flow structures.

\subsection{Kolmogorov-Zakharov spectra}
\label{sec:kz-spectra}

It is interesting to see that the kinetic equations in the
$f^3$-regime and the $f^2$-regime, Eqs.~\eqref{eq:dense} and
\eqref{eq:dilute}, respectively, accommodate more non-trivial steady
solutions than the thermal distribution.  Such solutions of the
power-law form are commonly called the Kolmogorov-Zakharov (KZ)
spectra, which are generally power-law spectra characterized by
exponents~\cite{kolmogorov1941local,zakharov1967weak,zakharov1972collapse}.
For the KZ solutions the collision integral becomes zero not due to
the detailed balance.  The gain and the loss from the higher energy
region and the lower energy region cancel out, which makes the
collision integral vanishing.  Although it is already an established
method, it would be instructive to make a quick review of the Zakharov
conformal transformation that is a mathematical trick to reverse the
higher/lower energy regions in a specific conformal way.

For the moment, as the following discussions can be applied to both
the $f^3$-regime and the $f^2$-regime, let us write the collision
integral in a generic form as
\begin{equation}
    Q = \int_{\energy_2,\energy_3,\energy_4} \!\!
        \delta(\energy_1 \!+\! \energy_2 \!-\! \energy_3 \!-\! \energy_4)
        \energy_{\min}^\alpha
        F(\energy_1, \energy_2, \energy_3, \energy_4)\;.
  \label{eq:Qf_02}
\end{equation}
where $F=[(1+f_1)(1+f_2)f_3 f_4-f_1 f_2(1+f_3)(1+f_4)]$ in the full
quantum case, $F=(f_1 f_3 f_4+f_2 f_3 f_4-f_1 f_2 f_3-f_1 f_2 f_4)$ in
the $f^3$-regime, and $F=(f_3 f_4-f_1 f_2)$ in the $f^2$-regime.

\begin{figure}
  \centering
  \includegraphics[width=0.3\textwidth]{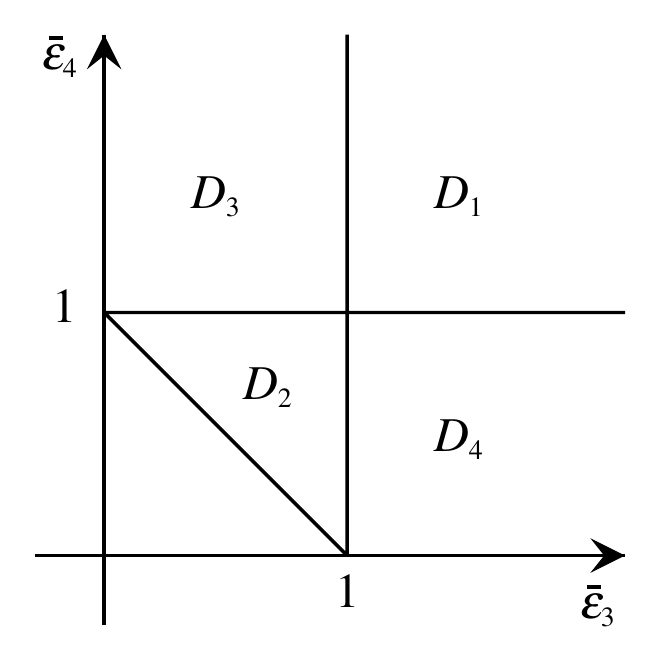}
  \caption{Four integration domains $D_1$, $D_2$, $D_3$, and $D_4$ in
    $\benergy_3$-$\benergy_4$ space.}
  \label{fig:z-transformation-domain}
\end{figure}

Because of the delta function, the $\energy_2$ integration is easily
done to substitute $\energy_2=\energy_3+\energy_4-\energy_1$.  Let us
now introduce dimensionless variables,
$\benergy_i \equiv \energy_i/\energy_1$, and then the collision
integral reads,
\begin{equation}
  Q = \energy_1^{2+\alpha} \int_{\cup_{i=1}^4 D_i} d\benergy_3\,
  d\benergy_4\, \benergy_{\min}^\alpha\,
  \bar{F}(\benergy_2, \benergy_3, \benergy_4; \energy_1)\;,
\end{equation}
where $\benergy_2 = \benergy_3 + \benergy_4 - \benergy_1$ with
$\benergy_1 = 1$.  The dimensionless integrand is defined as
$\bar{F}(\benergy_2, \benergy_3, \benergy_4; \energy_1)
= F(\energy_1, \energy_1\benergy_2, \energy_1\benergy_3, \energy_1\benergy_4)$.
The integration region with respect to $\benergy_3$ and $\benergy_4$
is split into four domains
$D_i \equiv \{(\benergy_3, \benergy_4)\,;\,\benergy_{\rm min} = \benergy_i\}$
($i=1,\dots,4$) as shown in Fig.~\ref{fig:z-transformation-domain}.
We note that the region for $\benergy_3+\benergy_4<1$ is excluded
because this region does not meet the energy conservation.

The Zakharov transformation is a conformal mapping among the domains
$D_i$.  We shall pick up one example.  For the integration over $D_2$,
we can change the integration variables as
$\benergy_3 \to \BEP{3}/\BEP{2}$ and $\benergy_4 \to \BEP{4}/\BEP{2}$,
where $\benergy_2 \to 1/\BEP{2}$ so that the energy conservation is
consistent also for $\BEP{i}$ with $\BEP{1}=1$.  Then, in the $D_2$
region, by definition,
$\benergy_{\min}^\alpha=\benergy_2^\alpha=\BEP{2}^{-\alpha}$, and a
factor $\BEP{2}^{-3}$ arises from the integration measure (note that
$\BEP{2}$ is a function of $\BEP{3}$ and $\BEP{4}$).  Most
importantly, the allowed region for $\BEP{3}$ and $\BEP{4}$ coincides
with $D_1$, i.e.\ $\BEP{3}>1$ and $\BEP{4}>1$.  Therefore, we have,
\begin{equation}
  \begin{split}
  & \int_{D_2} d\benergy_3\, d\benergy_4\, \benergy_{\min}^\alpha\,
  \bar{F}(\benergy_2,\benergy_3,\benergy_4;\energy_1) \\
  & \quad= \int_{D_1} d\BEP{3}\,d\BEP{4}\, \BEP{2}^{-\alpha-3}\,
  \bar{F}(\BEP{1}/\BEP{2},\BEP{3}/\BEP{2},\BEP{4}/\BEP{2};\energy_1)\;,
  \end{split}
  \label{eq:D21}
\end{equation}
where we inserted $\BEP{1}$ (instead of the unity) for a symmetric
representation.  Similarly, we can change the variables according to
respective regions of $D_i$ as summarized below:
\begin{center}
  \begin{tabular}{c|c}
    Domains & $(\benergy_2, \benergy_3, \benergy_4)$ \\ \hline
    $D_2\to D_1$ & $(\BEP{1}/\BEP{2}, \BEP{3}/\BEP{2}, \BEP{4}/\BEP{2})$ \\
    $D_3\to D_1$ & $(\BEP{4}/\BEP{3}, \BEP{1}/\BEP{3}, \BEP{2}/\BEP{3})$ \\
    $D_4\to D_1$ & $(\BEP{3}/\BEP{4}, \BEP{2}/\BEP{4}, \BEP{1}/\BEP{4})$
  \end{tabular}
\end{center}
The top on the table is the transformation as seen in
Eq.~\eqref{eq:D21}.  The second and the third transformations
introduce $\BEP{3}$ and $\BEP{4}$ in such a way that the allowed
region for them becomes $D_1$.  In this way, adding the original
integration in the $D_1$ domain (for which,
$\benergy_{\rm min}=\benergy_1=1$ and $\BEP{1}^{-\alpha-3}=1$ can be
safely inserted), we can find the collision integral transformed in
the following way in the $D_1$ domain only:
\begin{align}
  \begin{split}
    Q &= \energy_1^{2+\alpha}\int_{D_1} d\BEP{3}\, d\BEP{4} \\
    & \qquad\times \Bigl[
      \BEP{1}^{-\alpha-3}\bar{F}(\BEP{2}/\BEP{1},\BEP{3}/\BEP{1},\BEP{4}/\BEP{1};\energy_1) \\
    & \qquad\;\, + \BEP{2}^{-\alpha-3}\bar{F}(\BEP{1}/\BEP{2},\BEP{3}/\BEP{2},\BEP{4}/\BEP{2};\energy_1) \\
    & \qquad\;\, + \BEP{3}^{-\alpha-3}\bar{F}(\BEP{4}/\BEP{3},\BEP{1}/\BEP{3},\BEP{2}/\BEP{3};\energy_1) \\
    & \qquad\;\, + \BEP{4}^{-\alpha-3}\bar{F}(\BEP{3}/\BEP{4},\BEP{2}/\BEP{4},\BEP{1}/\BEP{4};\energy_1) \Bigr]\;.
  \end{split}
  \label{eq:Z-transformed-integral}
\end{align}
This expression is valid for any form of distribution function and its
functional $F$ as far as the original integration in each domain $D_i$
is convergent.

Now we shall find the KZ solutions of the form,
$f(\energy) \propto \energy^{-\gamma}$, with which the collision
integral vanishes.  The KZ solutions can exist if $F$ has the scaling
property,
\begin{equation}
  F(\energy_1/c, \energy_2/c, \energy_3/c, \energy_4/c)
  = c^{n\gamma} F(\energy_1, \energy_2, \energy_3, \energy_4)
\end{equation}
for an arbitrary number $c$ and an index $n$ fixed by $F$, and if $F$
has the symmetry under particle exchanges,
\begin{equation}
  F(\energy_1, \energy_2, \energy_3, \energy_4)
  = F(\energy_2, \energy_1, \energy_3, \energy_4)
  = -F(\energy_3, \energy_4, \energy_1, \energy_2)\;.
\end{equation}
Using the symmetries we can reorganize the integral as
\begin{align}
  \begin{split}
    & Q = \energy_1^{2+\alpha}\int_{D_1} d\BEP{3}\, d\BEP{4}\,
    F(\BEP{1}, \BEP{2}, \BEP{3}, \BEP{4}) \\
    &\times \bigl(\BEP{1}^{-\alpha-3+n\gamma} + \BEP{2}^{-\alpha-3+n\gamma}
    - \BEP{3}^{-\alpha-3+n\gamma} - \BEP{4}^{-\alpha-3+n\gamma} \bigr)\;.
  \end{split}
\end{align}
It is clear from the above expression that either
$-\alpha-3+n\gamma=0$ or $-\alpha-3+n\gamma=1$ makes $Q=0$, for which
the underlying mechanisms are to be identified as the particle number
conservation and the energy conservation,
respectively~\cite{Mehtar-Tani:2016bay}.  These relations lead to the
following KZ exponents,
\begin{equation}
  \gamma = \frac{\alpha+3}{n}\;,\qquad
  \gamma = \frac{\alpha+4}{n}\;.
\end{equation}
In the present work we will call these two solutions the KZ-I and the
KZ-II, respectively.

In the $f^3$-regime with Eq.~\eqref{eq:dense} the index is $n=3$, and
then more explicit forms of the KZ-I and the KZ-II solutions read,
\begin{equation}
  f_{\rm I}(\energy) = \energy^{-(\alpha+3)/3}\;,\qquad
  f_{\rm II}(\energy) = \energy^{-(\alpha+4)/3}\;.
  \label{eq:KZI}
\end{equation}
The KZ-I and KZ-II solutions correspond to the particle and the energy
flow, respectively, as we already mentioned above when we derived
$\gamma$.

In the $f^2$-regime with the index $n=2$, the KZ-I and the KZ-II
solutions are given, respectively, as
\begin{equation}
  f_{\rm I}(\energy) = \energy^{-(\alpha+3)/2}\;,\qquad
  f_{\rm II}(\energy) = \energy^{-(\alpha+4)/2}\;.
\end{equation}

\subsection{Self-similar evolving solution}
\label{sec:dilute}

We address a new type of solution in the $f^3$-regime and also the
$f^2$-regime, that is not a steady solution like the KZ spectra, but a
scaling solution having explicit time dependence.  We name it the
self-similar (SS) evolving solution.  The SS solution appears as
\begin{equation}
  f_{\rm SS}(\energy,t) = \Biggl[
    \frac{\energy^{-(2+\alpha-\dos)}}{I\,t + C} \Biggr]^{1/(n-1)}\;,
\end{equation}
where $C$ is a constant, and $\dos$ is an index characterizing the
density of states as $\rho(\energy) \propto \energy^{\dos}$.  The
coefficient $I$ is given by
\begin{equation}
  I = - \frac{(n-1)\,
    Q[f = \energy^{-(2 + \alpha - \dos)/(n-1)}](\energy_1 = 1)}
  {\rho(\energy_1 = 1)}\;.
\end{equation}
Here, in the above expression, a specific power-law functional form is
substituted for $f$, and after the $\energy_2$, $\energy_3$,
$\energy_4$ integrations in $Q$, we get rid of $\energy_1$ by taking
it to be the unity.  That is, $\rho(1)$ represents the coefficient
apart from the energy dependent part.

Let us explain how to find this SS solution in more details.  To this
end, we introduce an Ansatz, $f = A(t)\,\energy^{-\gamma}$.  Then,
the left-hand side of the kinetic equation becomes,
\begin{equation}
  \rho_1 \frac{\partial f(\energy_1)}{\partial t}
  = \rho(1)\, \dot{A}(t)\, \energy_1^{\dos-\gamma}\;.
\end{equation}
Again, here, we note that $\rho(1)$ represents the coefficient apart
from $\energy_1$ and the mass dimension is not skewed up.  The
right-hand side is,
\begin{align}
  Q[f](\energy_1) &= \energy_1^{2+\alpha}\int d\benergy_3\,
  d\benergy_4\, \benergy_{\min}^\alpha\,
  F(\energy_1, \energy_2, \energy_3, \energy_4) \notag\\
  &= Q[\energy^{-\gamma}](1)\, A(t)^n\,\energy_1^{2+\alpha-n\gamma}\;.
\end{align}
By equating above expressions, we readily find the following
choice of $\gamma$ and $A(t)$ is sufficient:
\begin{align}
  \gamma &= \frac{2 +\alpha-\dos}{n-1}\;, \\
  A(t) &= (I\,t + C)^{-1/(n-1)}\;.
\end{align}

For special combinations of $\alpha$ and $\dos$, the SS solution is
reduced to the KZ or the RJ solutions.  In such cases, $\gamma$ is the
exponent of the KZ/RJ solutions and $I$ vanishes so that $A(t)$ is
constant.  Such combinations of $\alpha$ and $\dos$ corresponding to
reduced solutions are summarized in Table~\ref{tab:ss-degeneracy}.

\begin{table}
  \centering
  \begin{tabular}{|c|c|c|c|} \hline
    \multicolumn{2}{|c|}{$f^3$-regime} &
    \multicolumn{2}{|c|}{$f^2$-regime} \\ \hline\hline
    Reduced Solution & $\alpha$, $\dos$ &
    Reduced Solution & $\alpha$, $\dos$  \\ \hline
    SS\,,\; KZ-I    & $\dos = \frac{1}{3}\alpha$ &
    SS\,,\; KZ-I    & $\dos = \frac{\alpha+1}{2}$ \\
    SS\,,\; KZ-II   & $\dos = \frac{\alpha-2}{3}$ &
    SS\,,\; KZ-II   & $\dos = \frac{\alpha}{2}$ \\
    SS\,,\; RJ      & $\dos = \alpha$ & & \\ \hline
  \end{tabular}
  \caption{Combinations of $\alpha$ and $\dos$ corresponding to
    solutions reduced from self-similar solutions.}
  \label{tab:ss-degeneracy}
\end{table}

\subsection{Intersection of solutions}
\label{sec:full-theory-scaling-solution}

For the full quantum case including both $f^2$ and $f^3$ terms as in
Eq.~\eqref{eq:Q}, there is in general no scaling solution of the form,
$f = A(t)\,\energy^{-\gamma}$.  Nevertheless, for special values of the
indices, $\alpha$ and $\dos$, we find that an SS solution in the
$f^2$-regime and the $f^3$-regime becomes an analytically exact
solution for the full quantum case.  This is the case typically for
indices that allow for some scaling solutions in the $f^2$-regime and
the $f^3$-regime simultaneously.

Such special indices are summarized in Table~\ref{tab:sskz} together
with the solutions in the $f^2$-regime and the $f^3$-regime, as well
as physically unaccepted solutions.

On Table~\ref{tab:sskz} the first three are physically possible, for
which the SS solution in the $f^2$-regime intersects with the
power-law solutions in the $f^3$-regime, so that the solution can
satisfy the full kinetic equation.  For $\dos<0$, that is unlikely in
physical systems, the SS solution in the $f^3$-regime intersects with
the power-law solutions in the $f^2$-regime, as listed on the 4th and
5th lines in Table~\ref{tab:sskz}.  Logically speaking, there are
combinations of indices that make the KZ solutions possible in the
$f^2$-regime and the $f^3$-regime at the same time.  Then, however,
$\alpha<0$ as listed on the last three lines in Table~\ref{tab:sskz},
which leads to infrared divergent $Q$ and is not physically
acceptable.

\begin{table}
  \centering
  \begin{tabular}{|c|c|} \hline
  $f^3$-regime / $f^2$-regime & Indices\\ \hline\hline
  RJ / SS         & $\dos = \alpha+1$ \\
  KZ-I / SS       & $\dos = \frac{2\alpha+3}3$ \\
  KZ-II / SS      & $\dos = \frac{2\alpha+2}3$ \\
  SS / KZ-I       & $\dos = -1$ \\
  SS / KZ-II      & $\dos = -2$ \\
  RJ,KZ-II / KZ-I & $\alpha = -1$ \\
  RJ / KZ-II      & $\alpha = -2$ \\
  KZ-I / KZ-II    & $\alpha = -6$ \\ \hline
  \end{tabular}
  \caption{Special indices of $\alpha$ and $\dos$ where the scaling
    solution is exact for the full theory.  Solutions with $\dos<0$ or
    $\alpha<0$ are physically unacceptable but shown for completeness
    of the listing.}
  \label{tab:sskz}
\end{table}

\section{Numerical results and flow diagram}
\label{sec:flow}

We shall now proceed to the numerical calculations to solve the boson
Boltzmann equation.  Because it is one-dimensionally reduced, it is
quite easy to solve the boson Boltzmann equation even in a brute-force
numerical way, and moreover, there is no subtlety in implementing the
energy-momentum conservation on the phase-space grid.  We could have
shown many numerical results for whole time evolution with various
initial conditions, but such presentations would not be illuminating
to deepen our understanding on general dynamics out of equilibrium.
We will thus first develop a new analysis to extract the information
inspired by the (perturbative) RG study.  Then, after making a remark
on the convergence of the collision integral that limits the sensible
parameter range, we will show the flow diagrams and discuss the
physical meaning of the critical lines.

\subsection{Method}
\label{subsec:Method}

To restrict our consideration within finite dimensional parameter
space, we here introduce three parameters, $\beta$, $\gamma$, and
$\mu$, to parametrize the distribution function as follows:
\begin{equation}
  f(\energy;\beta,\gamma,\mu) = \biggl[
    \frac{1}{e^{\beta(\energy-\mu)}-1} \biggr]^\gamma\;.
  \label{eq:fitter1}
\end{equation}
This parametrized Ansatz can encompass all kinds of solutions as we
have seen so far.  The choice of $\gamma = 1$ makes $f$ reduced to the
standard Bose-Einstein distribution function.  It should be noted that
$\beta$ and $\mu$ then have the ordinary interpretation as the inverse
temperature and the chemical potential, respectively.  The limit of
$\gamma \to 0$ makes $f$ constant and then $\beta$ and $\mu$ are
completely irrelevant.

The $f^3$-regime for $f \gg 1$ is realized in the low energy region
where $\beta(\energy-\mu) \ll 1$.  In this region $f$ has the
asymptotic form given by
$f(\energy) \approx \beta^{-\gamma} (\energy-\mu)^{-\gamma}$, which is
compatible to the RJ/KZ/SS solutions in the $f^3$-regime if
$\mu = 0$.  In the $f^2$-regime corresponding to the high energy
region of $f$, the function correctly reproduces the Maxwell-Boltzmann
distribution, $f(\energy) \approx e^{-\gamma\beta(\energy-\mu)}$.

\begin{figure}
  \centering
  \includegraphics[width=0.45\textwidth]{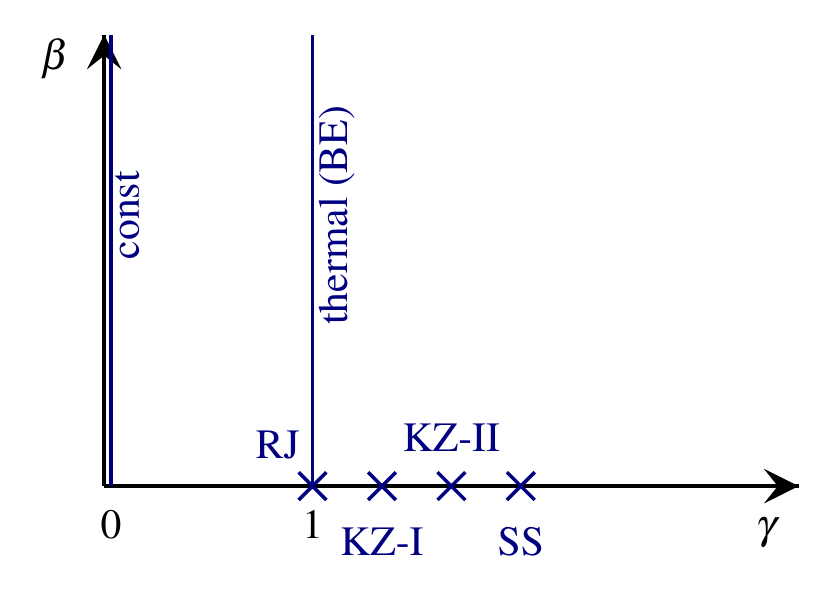}
  \caption{Schematic picture of fixed point solutions in our parameter
    space spanned by $\beta$ and $\gamma$.  The $\gamma = 0$ line and
    $\gamma = 1$ line correspond to the constant solutions and the
    thermal solutions, respectively.  The thermal distribution is
    further characterized by the temperature parameter $\beta$, while
    $\beta$ is indefinite for the constant solution.}
  \label{fig:fixed-points-in-parameter-space}
\end{figure}

If we are interested in the possibility of the Bose-Einstein
condensation, we should deal with non-zero $\mu$, but we can set
$\mu=0$ for the present work.  Then, all the fixed points as discussed
so far are located on $\beta$-$\gamma$ plane as schematically
illustrated in Fig.~\ref{fig:fixed-points-in-parameter-space}.

Now we are interested in the time evolution of $\beta$ and $\gamma$
on top of these fixed points.  Strictly speaking, the full time
evolution of $f$ cannot be completely captured by only two parameters
in Eq.~\eqref{eq:fitter1}.  In this sense, our approach somehow shares
the truncation scheme with the perturbative RG flow analysis, for
which the functional space is restricted to the one described by a
finite number of couplings.

What we are doing is the following.  We will compute the time
derivatives of $\beta$ and $\gamma$ at each point ($\beta$,$\gamma$)
to show the vector that represents the flow direction at that point.
We also implicitly assume looking at a narrow energy window around
$\energy^\ast$.  Then, the local shape of the distribution function is
well approximated by its local value and derivative, $f(\energy^\ast)$
and $\partial_{\lnE}f(\energy^\ast)$, with $\lnE=\ln\benergy$.  Now,
to shorten the notation, let us denote the energy differentiated $f$
as $f'(\energy) \equiv \partial_{\lnE} f(\energy)$.  For an
infinitesimal time increase, the time evolution of the distribution
function is also approximated by local quantities differentiated with
respect to the time, i.e.\
$\dot{f}(\energy^\ast)$ and $\dot{f}'(\energy^\ast)$.  We can
numerically obtain these time derivatives from the collision integral.
We make a remark that it is convenient to employ the Zakharov
transformed expression~\eqref{eq:Z-transformed-integral} for the
numerical integration because it guarantees the exact numerical
cancellation in the collision integral at the KZ fixed points.

Then, we need to evaluate $\dot{\beta}$ and $\dot{\gamma}$ from
$\dot{f}$.  The idea is that $\dot{\beta}$ and $\dot{\gamma}$ should
best reproduce the shape of (non-truncated) $\dot{f}$.  Up to the
first order in terms of the energy derivatives, the following matrix
equation must hold:
\begin{equation}
  \begin{pmatrix}
    \frac{\partial f(\energy^\ast; \beta, \gamma)}{\partial\beta} &
    \frac{\partial f(\energy^\ast; \beta, \gamma)}{\partial\gamma} \\
    \frac{\partial^2 f'(\energy^\ast; \beta, \gamma)}{\partial\beta} &
    \frac{\partial^2 f'(\energy^\ast; \beta, \gamma)}{\partial\gamma}
  \end{pmatrix}
  \begin{pmatrix}
    \dot{\beta} \\
    \dot{\gamma}
  \end{pmatrix} =
    \begin{pmatrix}
      \dot{f}(\energy^\ast) \\
      \dot{f}'(\energy^\ast)
    \end{pmatrix}\;.
  \label{eq:chain-rule-eq-for-parameter-flow}
\end{equation}
Here, we notice that this matrix form can be trivially extended for a
larger number of parameters by taking account of the higher order
energy derivatives.

The vector field,
$(\dot{\beta}(\beta,\gamma),\dot{\gamma}(\beta,\gamma))$, makes our
``flow'' diagram associated with the kinetic equation.  More details
on our concrete numerical procedures are explained in
Appendix~\ref{app:chain-rule}.  The flow diagram enables us to discuss
the global structure of the time evolution of the kinetic equation as
discussed in Sec.~\ref{subsec:Flow-diagrams}.

\subsection{Convergence of collision integral}
\label{subsec:Convergence}

Before we turn to discuss our numerical results, we will briefly
discuss the validity region of our analysis in parameter space.  Since
the original collision integral has some parameter space without
absolute convergence, the flow results there should not be trustable
even though the behavior seems non-singular.  Thus, it is
important to quantify the convergence condition for the collision
integral.

For the numerical integration of the collision integral, we adopt the
expression~\eqref{eq:Z-transformed-integral} after the Zakharov
transformation, which chooses a specific combination of the
integration domains.  Even when a regular output results from our
analysis, therefore, it does not necessarily describe physically
sensible behavior unless the original collision integral converges.

Since $e^{\beta\energy}-1\le 1$ for $\beta\le 0$, the fitting function
is no longer positive definite then.  So we should first require
$\beta > 0$.  To guarantee the ultraviolet convergence, we should next
require $\gamma > 0$.  Let us see below how the infrared properties
further constrain allowed $\gamma$.  As a matter of fact, because the
distribution function has an approximate power form
$f\sim\energy^{-\gamma}$ in the infrared region, a larger $\gamma$
would suffer a stronger infrared singularity.  Here, we note that,
though we could analyze the collision integral~\eqref{eq:Qf_02}
directly, it would be easier to start with the
expression~\eqref{eq:Z-transformed-integral} after the Zakharov
transformation, with which original ``infrared'' divergences in
Eq.~\eqref{eq:Qf_02} are transformed into ``ultraviolet'' divergences
in Eq.~\eqref{eq:Z-transformed-integral}.  It should be noticed that
we need to evaluate each term in Eq.~\eqref{eq:Z-transformed-integral}
separately to check the absolute convergence of the original collision
integral.  There are two possible situations having divergences.  One
is the case that either $\benergy_3$ or $\benergy_4$ gets infinity.
The other possibility is the case that both $\benergy_3$ and
$\benergy_4$ get infinity.  We obtain two different power-law
behaviors with respect to an ultraviolet cut-off $\Lambda$
corresponding to these two situations.  By reading the exponents of
the power-law behaviors, we can quantitatively identify the validity
regions.

In the $f^2$-regime, the collision integral, whose explicit form
appears in Eq.~\eqref{eq:dilute}, yields
$\sim\Lambda^{\gamma-\alpha-1}$.  For the convergence, the power of
such $\Lambda$ dependence should be non-positive, that is,
\begin{equation}
  \gamma \leq \alpha + 1\;.
\end{equation}
In the $f^3$-regime, when $\benergy_4\to\Lambda$ with $\benergy_3$
being finite, the collision integral \eqref{eq:dense} behaves as
$\sim\Lambda^{2\gamma-\alpha-2}$.  If both $\benergy_3$ and
$\benergy_4$ approach $\Lambda$, the collision integral is
$\sim\Lambda^{\gamma-\alpha-1}$.  As long as $\alpha>0$, the former
condition is stronger than the latter and $\gamma$ is constrained as
\begin{equation}
  \gamma \leq \frac{1}{2}\alpha + 1\;.
\end{equation}
For the full quantum case with $f^2$ and $f^3$ terms in the collision
integral, the convergence condition follows from the stronger one of
above two limiting cases.  For example, for $\alpha=2$, the condition
in the $f^2$-regime is $\gamma\le 3$ and that in the $f^3$-regime is
$\gamma \le 2$, so that the condition in the full dynamics is given by
the latter ($\gamma\le 2$) that is stronger than the former
($\gamma\le 3$).

\subsection{Flow diagrams}
\label{subsec:Flow-diagrams}

\begin{figure}
  \centering
  \includegraphics[width=0.48\textwidth]{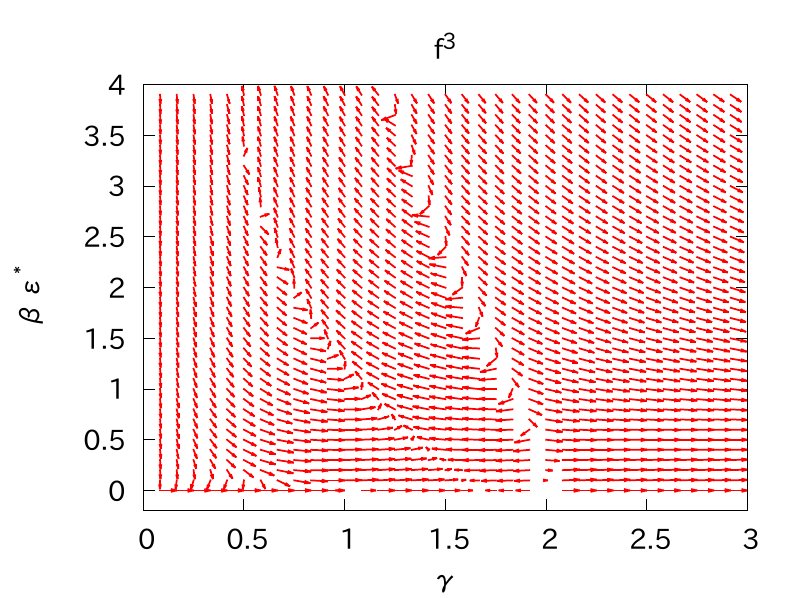}
  \caption{The flow diagram for the $f^3$-regime.}
  \label{fig:flow2}
\end{figure}
\begin{figure}
  \centering
  \includegraphics[width=0.48\textwidth]{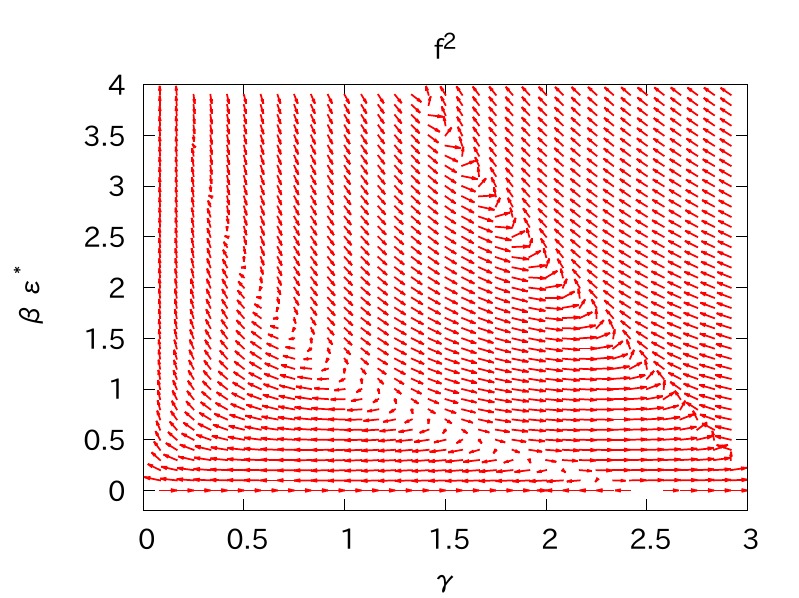}
  \caption{The flow diagram for the $f^2$-regime.}
  \label{fig:flow3}
\end{figure}
\begin{figure}
  \centering
  \includegraphics[width=0.48\textwidth]{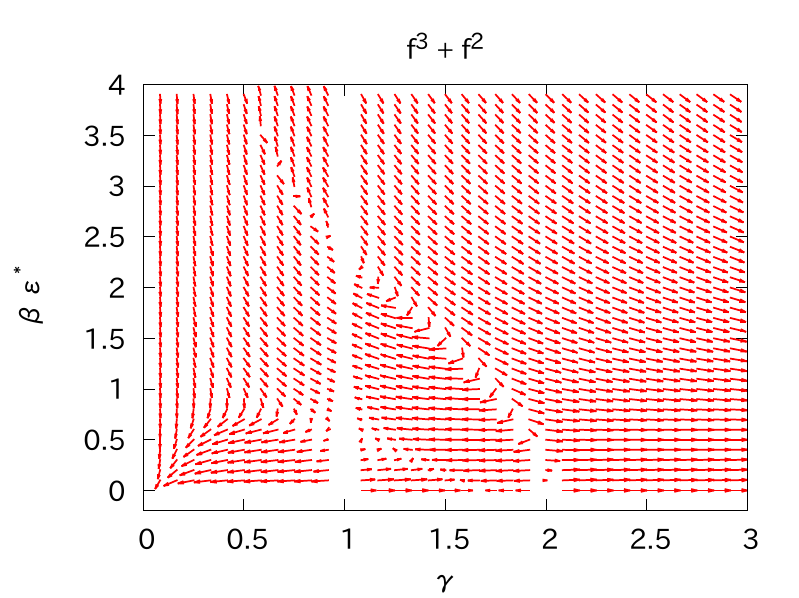}
  \caption{The flow diagram for the full collision integral including
  the $f^2$ and $f^3$ terms.}
  \label{fig:flow1}
\end{figure}

Our main results of the flow diagrams are summarized in
Figs.~\ref{fig:flow2}, \ref{fig:flow3}, and \ref{fig:flow1} for
the $f^3$-regime with Eq.~\eqref{eq:dense},
the $f^2$-regime with Eq.~\eqref{eq:dilute},
and the full quantum case with Eq.~\eqref{eq:Q},
respectively.  In the present work we chose $\alpha=\dos=2$ (there is
no particular reason for this choice).  To draw figures we took
$\gamma$ for the horizontal axis and $\beta\energy^\ast$ for the
vertical axis;  in our Ansatz, only a product of $\beta\energy^\ast$
appears.  Thus, to draw these figures, we changed $\beta$ with the
energy fixed at $\energy^\ast=1$.  For the graphical representation we
rescaled the length of the vector from
$l=\sqrt{ (\dot{\beta}\energy^\ast)^2 + (\dot{\gamma}^\ast)^2 }$ to
$a \tanh[l/N(\beta,\gamma)]$ where we chose $a=0.08$ and
$N(\beta,\gamma)=f^\ast(f^\ast+1)$, ${f^\ast}^2$, and
$f^\ast$ for the full quantum case, the $f^3$-regime, and the
$f^2$-regime, respectively, with
$f^\ast\equiv f(\energy^\ast;\beta,\gamma)$ defined.  We note that
Fig.~\ref{fig:flow1} does not show points
for $\beta\energy^\ast=10^{-6}$ and $0\le \gamma < 1$ because of slow
convergence of the numerical calculation.

On these flow diagrams we anticipate that fixed points should manifest
themselves as points where the flows stop.  In fact, we can easily see
such points on these flow diagrams in accord with discussions and a
schematic picture given in Sec.~\ref{sec:fixedpoints}.

In the $f^3$-regime, we can locate three fixed points precisely
corresponding to power-law solutions on the horizontal axis (on
$\beta\energy^\ast = 0$): the RJ ($\gamma = 1$), the KZ-I
($\gamma = 5/3$), and the KZ-II ($\gamma = 2$) solutions on
Fig.~\ref{fig:flow2}.  In the $f^2$-regime, on Fig.~\ref{fig:flow3},
there appear two fixed points:  the KZ-I ($\gamma = 5/2$) and the
KZ-II ($\gamma = 3$) solutions.

For the full quantum case, that is of our main interest, we can find
the Bose-Einstein solutions at $\gamma = 1$ with various temperatures
$\beta\energy^\ast$ along the vertical line on Fig.~\ref{fig:flow1}.
We can see the power-law solutions of the $f^3$-regime near
$\gamma=5/3$ and $2$ also in this case with the full quantum terms.
This is because the occupation number $f$ becomes large near the
horizontal axis $\beta\energy^\ast \approx 0$, so that the full
quantum equation~\eqref{eq:Q} is effectively reduced to that of the
$f^3$-regime~\eqref{eq:dense}.

The most remarkable feature on these flow diagrams is not the
manifestation of the fixed points, but the flow pattern around the
fixed points.  Actually we see that the flow directions form several
distinct regions, with which the whole parameter space is split into
different ``phases'' that is again reminiscent of the perturbative RG
flow diagram.  Interestingly, for each power-law solution, as seen in
Figs.~\ref{fig:flow2} and \ref{fig:flow3}, one ``line'' with rapid
change of the flow directions is always attached to one fixed point,
which defines the borders of phases.  Moreover, in the full quantum
case as in Fig.~\ref{fig:flow1}, those lines are crossed with the
thermal line at $\gamma=1$ to shape a more complicated phase
structure.  In the next subsection we discuss these lines more
closely.

\subsection{Flow lines and phase boundaries}
\label{subsec:Flow-lines-diagrams}

\begin{figure}
  \centering
  \includegraphics[width=0.48\textwidth]{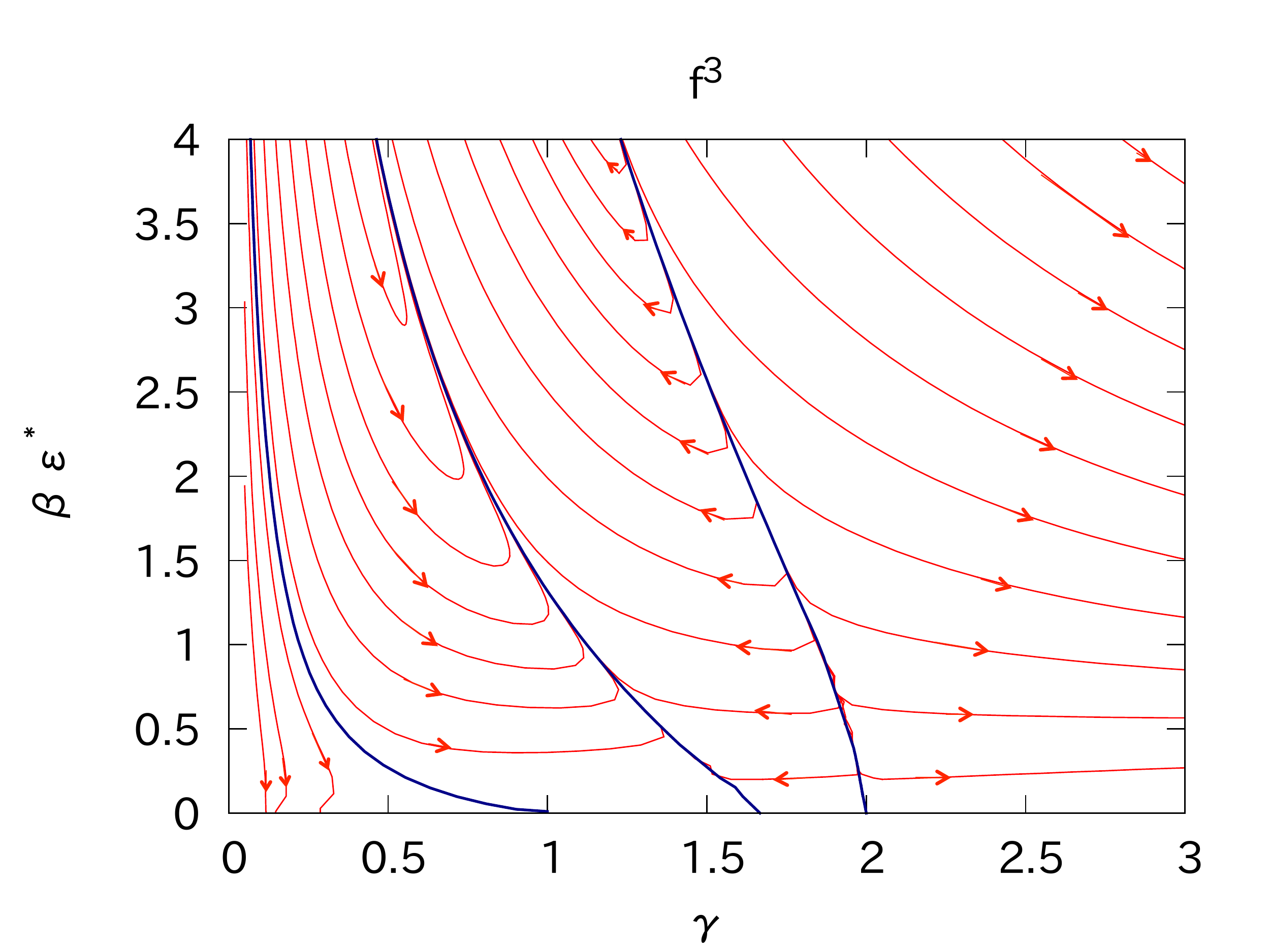}
  \caption{Flow lines on the flow diagram shown for the same
    setup as Fig.~\ref{fig:flow2}.}
  \label{fig:flowline2}
\end{figure}
\begin{figure}
  \centering
  \includegraphics[width=0.48\textwidth]{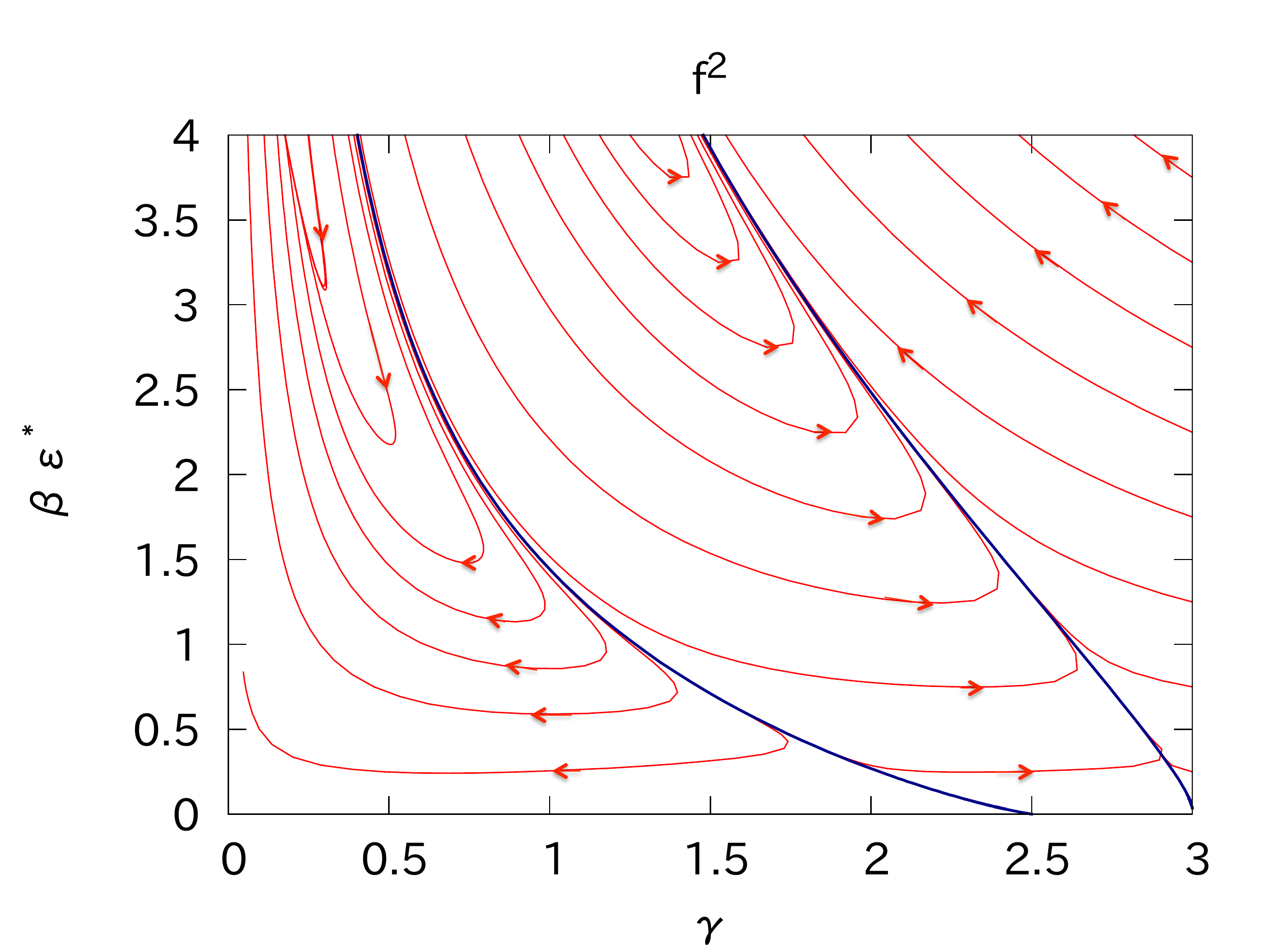}
  \caption{Flow lines on the flow diagram shown for the same
    setup as Fig.~\ref{fig:flow3}.}
  \label{fig:flowline3}
\end{figure}
\begin{figure}
  \centering
  \includegraphics[width=0.48\textwidth]{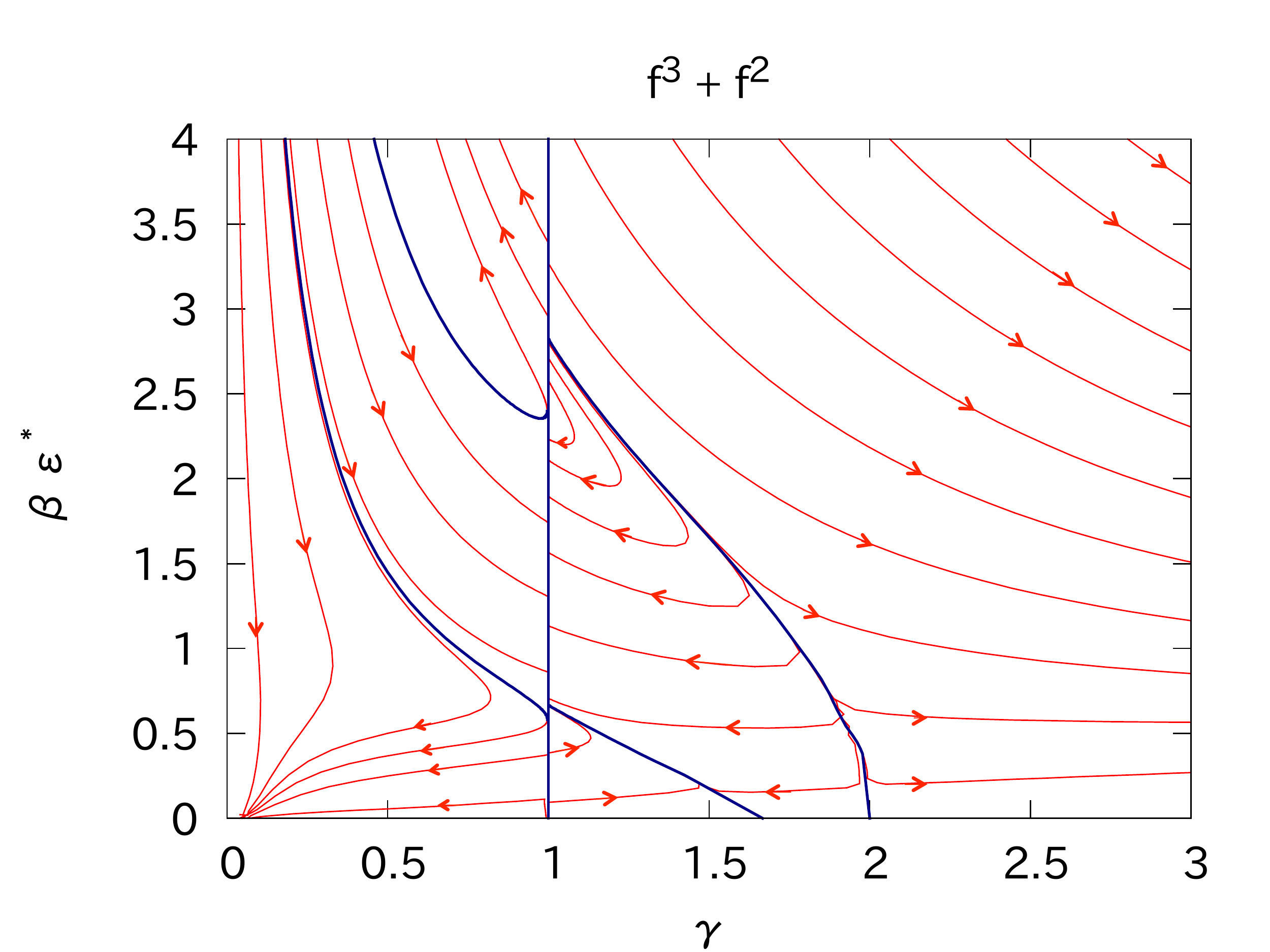}
  \caption{Flow lines on the flow diagram shown for the same
    setup as Fig.~\ref{fig:flow1}.}
  \label{fig:flowline1}
\end{figure}

To make the phase structure visible more prominently, and to
understand how the fixed points are connected by phase boundaries, it
would be very useful to consider ``flow lines'' on the diagrams as
shown in Figs.~\ref{fig:flowline2}-\ref{fig:flowline1}.  Instead of
the vector field $(\dot{\beta}\energy^\ast, \dot\gamma)$ as discussed
in Sec.~\ref{subsec:Flow-diagrams}, we would pay our attention to
their integral curves, which we call the ``flow lines'' throughout
this paper.  We can identify the flow lines by solving the following
set of equations using the Runge-Kutta 3/8-rule, i.e.\ one of the
$4^{\rm th}$ order methods:
\begin{equation}
  \frac{d}{ds} \begin{pmatrix}
    \beta(s)\\
    \gamma(s)
  \end{pmatrix} = \begin{pmatrix}
    \dot\beta(\beta(s),\gamma(s)) \\
    \dot\gamma(\beta(s),\gamma(s))
  \end{pmatrix}\;,
\end{equation}
where $s$ is a parameter along the curve.  In
Figs.~\ref{fig:flowline2}-\ref{fig:flowline1} we show flow lines by
red lines together with supplementary arrows indicating the flow
directions.  It should be noted that initial points
$(\beta(0), \gamma(0))$ are chosen by hand arbitrarily.

Although the precise locations of respective flow lines are
irrelevant, it is physically meaningful where the phase ``boundaries''
appear, which are shown by blue lines on
Figs.~\ref{fig:flowline2}-\ref{fig:flowline1}.  The most trivially
understandable phase boundary is the thermal line at $\gamma = 1$ in
the full quantum case in Fig.~\ref{fig:flowline1}.  It is also clear
that the flow line starting from the KZ-I point is an
\textit{attractive line} in the full quantum case as well as in the
$f^3$-regime in Fig.~\ref{fig:flowline2};  surrounding flow lines
around the boundary run in the direction approaching the attractive
line.  In contrast, the flow line starting from the KZ-II point is a
\textit{repulsive line}, i.e.\ flow lines branch out from this
unstable repulsive line.  In the $f^2$-regime, similarly, the boundary
starting from the KZ-I (and KZ-II) point is a repulsive (and
attractive, respectively) line.

We point out that there is another kind of phase boundary in the
$0 < \gamma < 1$ region in the full quantum case as noticed in
Fig.~\ref{fig:flowline1}.  This $\gamma$ region is further divided by
two phase boundaries into three distinct phases according to the final
destinations of the flow.  The destination in the small
$\beta\energy^\ast$ phase is the origin on the diagram which
corresponds to a constant solution.  The destination in the middle
$\beta\energy^\ast$ phase is the thermal line.  In the large
$\beta\energy^\ast$ phase, the flow tends to go to
$\beta\energy^\ast\to\infty$.  We note that these structures share
features seen also in the $f^3$-regime in Fig.~\ref{fig:flowline2}.


\subsection{More discussions}

Here we take a closer look at the flow diagrams and discuss the
implications to the real-time dynamics.  We shall consider only the
full quantum case in Fig.~\ref{fig:flow1} in this subsection, but the
generalization of the interpretations for Figs.~\ref{fig:flow2} and
\ref{fig:flow3} is straightforward.

In the following discussion let us regard $\energy^\ast$ as a
changeable variable rather than $\beta$.  This implies that, for a
given $\gamma$, we can obtain qualitative information on the time
evolution of the whole distribution function by looking at the flow
pattern on the diagram.

\begin{figure}
  \includegraphics[scale=0.48]{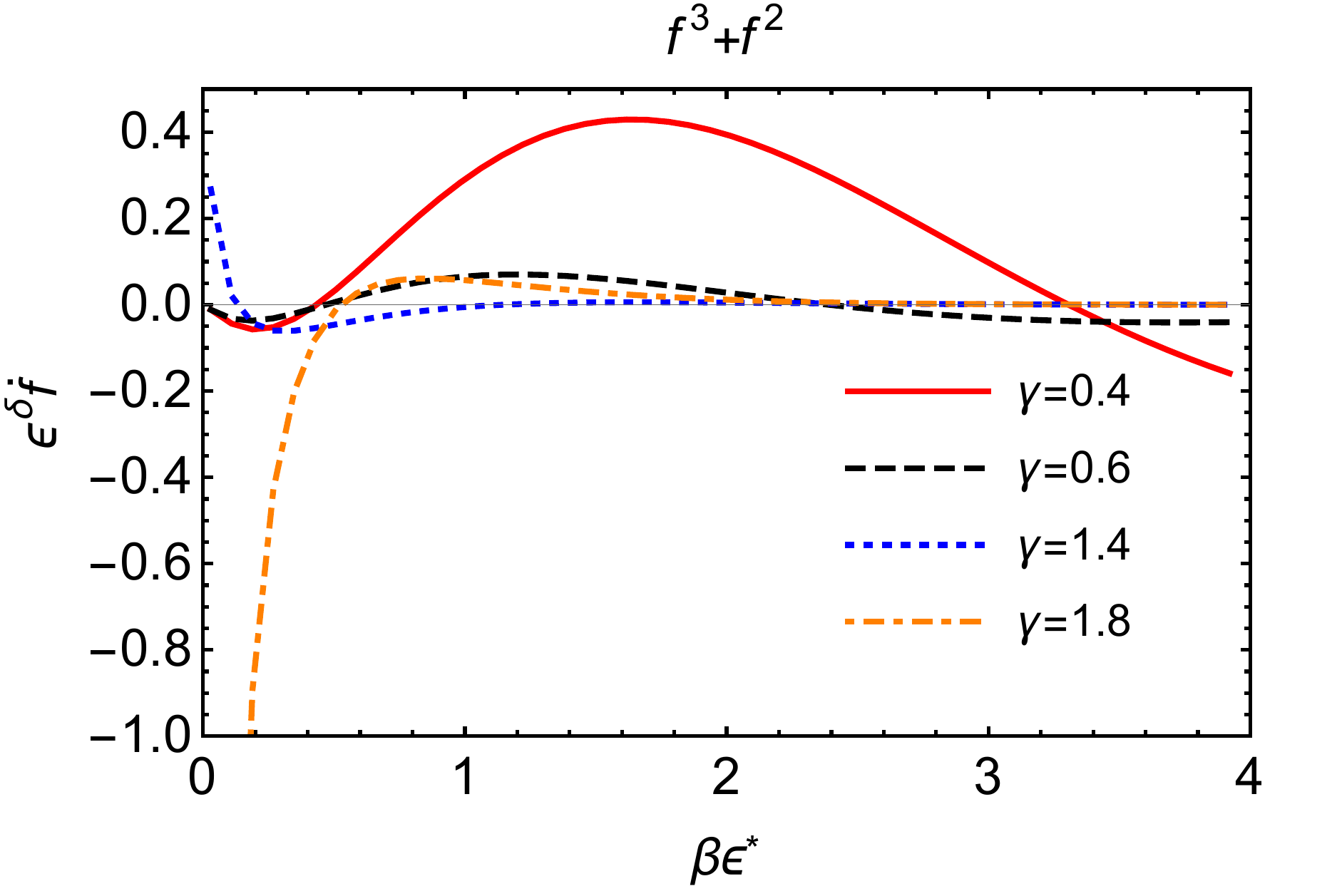} 
  \caption{Collision integral,
    $Q[f]=\energy^{\dos} \dot{f}$,
    for $\alpha=2$.  The red solid,
    the black dashed, the blue dotted, and the yellow dot-dashed lines
    stand for $\gamma=0.4$, $0.6$, $1.4$, $1.8$, respectively.}
  \label{fig:Qf_dQf}
\end{figure}

In Fig.~\ref{fig:flow3} we notice that the vectors
$(\dot{\beta}\varepsilon^\ast,\dot{\gamma})$ have the opposite signs
in the left and the right sides from the line at $\gamma=1$.  To
discuss underlying physics for such different behavior, we divide the
flow diagram into three regions:  the region-I refers to $0<\gamma<1$,
the region-II refers to $1\le\gamma<(\alpha+3)/3$, and the region-III
refers to $(\alpha+3)/3\le\gamma<2$.  We should remember that, as
already mentioned in Sec.~\ref{subsec:Convergence}, $\gamma>2$ makes
the collision integral non-convergent, and thus, we should exclude
this $\gamma>2$ region from our consideration.  In order to explain
the flow diagram better, we make plots for the collision integral as a
function of $\beta\energy^\ast$ for various values of $\gamma$ in
Fig.~\ref{fig:Qf_dQf}.

For $0<\gamma<1$ (region-I), from Fig.~\ref{fig:Qf_dQf}, we see that
$\dot{f}$ at small and large energies is negative, and so $f$ in these
energy regions should decrease.  In the middle energy region $\dot{f}$
is positive and the system accumulates more particles in this middle
energy region.

Because the distribution function naturally flows toward
thermalization, we can anticipate that more and more particles will be
transferred from large $\beta\energy^\ast$ to small
$\beta\energy^\ast$ through interactions.  In fact, to satisfy the
particle number conservation and the energy conservation
simultaneously, a particle at very large $\beta\energy^\ast$ must
interact with a particle at very small $\beta\energy^\ast$, turning
into a positive contribution to a particle at middle
$\beta\energy^\ast$.  This is an intuitive explanation for the
observation that $\dot{f}$ is negative at small and large energies,
while it is positive in the middle energy region in
Fig.~\ref{fig:Qf_dQf}.

Now, let us take a turn back to the flow diagram of
Fig.~\ref{fig:flow1}.  For large $\beta\energy^\ast$, the flow is
directed from high energy to low energy, while the flow changes its
direction around $\beta\energy^\ast\approx \ln2\simeq0.69$.
As is clear in Fig.~\ref{fig:flow1}, for $\beta\energy^\ast < \ln 2$,
the direction of flow is almost parallel to the $\gamma$-axis, that
means the energy is hardly changed along the flow though the number of
soft particles increases.  Actually, $\beta\energy^\ast=\ln2$ is a
special point that makes our Ansatz~\eqref{eq:fitter1} as simple as
$f=1^\gamma=1$ regardless of $\gamma$.  Therefore, the sensitivity of
$\gamma$ becomes far stronger then and the flow should be almost
parallel to the $\gamma$-axis.

For $1\le \gamma<(\alpha+3)/3$ (region-II), the collision integral is
positive for small $\beta\energy^\ast$ only as shown by the blue
dotted line in Fig.~\ref{fig:Qf_dQf}.  We argue that the system tends
to reach the KZ-I solution rather than moving straight to thermal
equilibrium, which accounts for the observation that the system will
accumulate more soft particles only.  Turning back to the flow diagram
in Fig.~\ref{fig:flow1}, we see that the flow is again approximately
parallel to the $\gamma$-axis in small $\beta\energy^\ast$ region.
Along the flow $\gamma$ gets larger as the time goes, and this means
that the number of soft particles increases.  Then, for larger
$\beta\energy^\ast$ the flow pattern changes and becomes similar to
the one in the region-I with small $\beta\energy^\ast$.  For further
larger $\beta\energy^\ast$, the flow behavior is just the same as that
in the region-I.

For $(\alpha+3)/3<\gamma<2$ (region-III), the flow pattern looks
similar to that in the region-I at middle energy, and the underlying
physic is similar.  In the small $\beta\energy^\ast$ region the
system is over-populated especially near the KZ-II solution.  Then,
since we have not considered a possible condensate (with $\mu=0$
entirely) in our analysis, the particle number conservation does not
allow soft particles to increase.  As a results, the system will
accumulate more particles in the middle energy region only.  That is
the reason why the region-III exhibits some similarity to the
region-I.



\section{Conclusion}
\label{sec:conclusion}

In this work we classified various non-trivial fixed points in space
of the distribution function described by the boson Boltzmann
equation.  We proposed a new graphical way to analyze the dynamical
structures, i.e.\ the \textit{flow diagrams} and
the \textit{phase diagrams} resulting from the boson Boltzmann
equation.  For bosonic systems the most well-known thermal fixed point
is the Bose-Einstein (BE) distribution, that is a solution satisfying
the detailed balance with quantum terms in the collision integral.  In
the lowest order $2\leftrightarrow 2$ scattering, the full collision
integral contains terms involving both $f^2$ and $f^3$.  In the dense
limit which we call the $f^3$-regime in this work, the collision
integral keeps only the $f^3$ terms, leading to the Rayleigh-Jeans
(RJ) approximated form of the thermal distribution function.  In the
dilute limit or the $f^2$-regime, on the other hand, the collision
integral is truncated only with the $f^2$ terms and the thermal
distribution is approximated by the Maxwell-Boltzmann (MB) form.  In
the $f^3$/$f^2$ regimes there are additional non-trivial solutions of
the boson Boltzmann equation:  Two Kolmogorov-Zakharov spectra,
namely, KZ-I and KZ-II, are non-trivial power-law fixed points
corresponding to the particle and the energy cascade, respectively.
Furthermore, we addressed a new type of power-law fixed point:  We
found the self-similar (SS) evolving solutions whose overall factor
has explicit time dependence.  Interestingly, even in the full quantum
case with both the $f^2$ and the $f^3$ terms, there can exist
power-law solutions at the interaction of those non-trivial fixed
points in the $f^2$/$f^3$ regimes.

We postulated a parametrization of the distribution function so that
thermal fixed points and other non-trivial power-law fixed points are
interpolated and mapped into parameter space spanned by
$(\beta, \gamma)$.  The time evolution of the parameters,
$(\dot\beta, \dot\gamma)$, indicates the directions of infinitesimal
(quasi-static) temporal changes from the parametrized initial
condition, and were numerically obtained from the local time evolution
around an energy window $\energy^\ast$.  We constructed the flow
diagrams by plotting the vector field, $(\dot\beta, \dot\gamma)$, in
the two-dimensional $(\beta, \gamma)$ plane.  We observed a clear
manifestation of fixed points corresponding to the thermal
distribution, the KZ-I, and the KZ-II solutions on the flow diagram.
Besides, we noticed characteristic flow patterns around these fixed
points.  The whole parameter space is then split into several distinct
\textit{phases} bounded by \textit{critical lines}, which is
intuitively understood in analogy to the perturbative RG flow
diagrams.  To investigate more clear relations of fixed points and
critical lines, we numerically identified the \textit{flow lines} on
the flow diagrams.  Our concrete demonstration shows that the
clarification of the fixed points, the critical lines, and the flow
lines should provide us with useful information on the thermalization
processes including transient behavior at intermediate stages.  In the
full quantum case the flow lines become far complicated with the
thermal (BE) critical line that crosses flow lines.  The flow diagram
also tells us transparently that the critical lines starting from the
KZ-I and the KZ-II points are directed with attractive and repulsive
behavior, respectively.


As we emphasized, our proposed tools of the RG-like flow diagram offer
us an intuitive access to investigate the dynamics out of
equilibrium.  In the present study one might have thought that our
arguments may rely on special setups of the boson Boltzmann equation,
the $2\leftrightarrow 2$ scattering, the Ansatz of the distribution
function with $(\beta, \gamma)$, but none of them is a crucial
limitation.  Because we aimed to exemplify how useful our new analysis
is, we employed the simplest setup in the present work.  We can almost
trivially extend our current treatment to more general systems.  With
sufficient computational resources, in principle, one can numerically
solve the kinetic equation in full phase space, including inelastic
and higher-order collisions terms.  Such improvements would
quantitatively affect $\dot{f}$, that can be translated into
$(\dot{\beta}, \dot{\gamma})$.  Rather, depending on the problem of
our interest, we need to consider some other types of parametrization
of the distribution function.  For example, in this work, we reported
a new solution of the boson Boltzmann equation called the SS solution,
but the Ansatz we adopted was not suitable for confirming it
numerically.  We tested many other functional forms and, in fact, some
of them were capable of seeing the SS solution properties, but they
did not have good resolution for other fixed points.  The optimal
parametrization awaits to be found.  Although we did not pay much
attention to the SS solution in this paper, its physical implication
must be an intriguing problem in the analytical aspect of the
Boltzmann equation, which deserves future investigations.

\acknowledgments
The authors thank Fran\c{c}ois~Gelis, Yoshimasa~Hidaka,
Jinfeng~Liao, Yacine~Mehtar-Tani for useful discussions and comments.
K.F.\ and K.M.\ were partially supported by JSPS KAKENHI Grant
No.\ 15H03652 and 15K13479.
S.P.\ was supported by the JSPS Postdoctoral Fellowship for Foreign
Researchers.

\appendix

\section{Derivation of Eqs.~\eqref{eq:Q} and \eqref{eq:S_min_01}}
\label{sec:appA}

In the relativistic case, $\energy=|\bp|$, the collision integral
$C[f]$ in Eq.~\eqref{eq:collision_01} can be written as,
\begin{equation}
  \begin{split}
    C[f] &= \frac{1}{2\energy_1} \int \prod_{i=2}^4
    \frac{p_i^2\,d^3 p_i}{(2\pi)^3 (2\energy_i)}\,
    \delta(\energy_1 + \energy_2 - \energy_3 - \energy_4) \\
    &\quad\times |\mathcal{M}_{12\to 34}|^2 \, F(f) \\
    &\quad\times \int d\Omega_2\,d\Omega_3\,d\Omega_4\,
    \delta^{(3)}(\bp_1 \!+\! \bp_2 \!-\! \bp_3 \!-\! \bp_4)\;,
    \end{split}
\end{equation}
where $d\Omega_i$ represents the angular part of the phase space
integration.  For the full quantum process, the interaction involves,
\begin{equation}
  F(f) = f_1 f_2 (1+f_3)(1+f_4) - (1+f_1)(1+f_2) f_3 f_4\;.
\end{equation}
We can explicitly carry out the angular integration with the delta
function constraint as follows,
\begin{align}
  & \int d\Omega_2\,d\Omega_3\,d\Omega_4\,
  \delta^{(3)}(\bp_1 + \bp_2 - \bp_3 - \bp_4) \notag\\
  =& \int \frac{d^3 z}{(2\pi)^3} d\Omega_2\,d\Omega_3\,d\Omega_4\;
  e^{i\boldsymbol{z}\cdot(\bp_1+\bp_2-\bp_3-\bp_4)} \notag\\
  =& \int\frac{z^2\,dz}{(2\pi)^3} \int d\Omega_z\,e^{i\boldsymbol{z}\cdot\bp_1}
  \; d\Omega_2\,e^{i\boldsymbol{z}\cdot\bp_2}\;
  d\Omega_3\,e^{-i\boldsymbol{z}\cdot\bp_3}\;
  d\Omega_4\,e^{-i\boldsymbol{z}\cdot\bp_4} \notag\\
  =& \frac{16\pi^{4}}{(2\pi)^3 |\bp_1||\bp_2||\bp_3||\bp_4|}
  \int\frac{dz}{z^2}\prod_{i=1}^4 \sin(z|\bp_i|) \notag\\
  =& \frac{\pi^{2}}{2 |\bp_1||\bp_2||\bp_3||\bp_4|}
  \min\{|\bp_i|\}\;.
  \label{eq:anguler_01}
\end{align}
Inserting the above to $C[f]$ yields,
\begin{equation}
  \begin{split}
    C[f] &= \frac{g^{2}}{2\energy_1|\bp_1|}\frac{\pi^{2}}{2}\int
    \prod_{i=2}^4 \frac{p_i\,dp_i}{(2\pi)^3 (2\energy_i)} \\
    &\qquad\times \delta(\energy_1+\energy_2-\energy_3-\energy_4)\,F(f)\,
    \min\{|\bp_i|\}\;,
  \end{split}
\end{equation}
where we set $|\mathcal{M}_{12\to 34}|^2=g^2$ assuming a simple
interaction like that in the $\phi^4$ theory.  Recalling that the
density of states is given as
\begin{equation}
  \rho(\energy) \equiv \int\frac{d^3 p}{(2\pi)^3}\,
  \delta(\energy-|\bp|)
  \propto \energy^2\;,
\end{equation}
we finally get,
\begin{equation}
  \rho_1\frac{\partial f}{\partial t} = g^2\int d\energy_2\,
  d\energy_3\,d\energy_4\,\delta(\energy_1+\energy_2-\energy_3-\energy_4)
  \,F(f)\,\energy_{\min}\;.
\end{equation}
We note that, in the above expression, we absorbed all irrelevant
constant factors into a redefinition of $g^2$.

In the same way we can derive the boson Boltzmann equation for the
non-relativistic case.  The difference is that the energy dispersion
relation is not $\energy=|\bp|$ but $\energy=|\bp|^2/(2m)$.  The
collision integral $C[f]$ then becomes,
\begin{equation}
  \begin{split}
    C[f] & = \bigl(\frac{1}{2m}\Bigr)^4 \int\prod_{i=2}^4
    \frac{p_i^2\,dp_i}{(2\pi)^3}\,
    \delta(\energy_1+\energy_2-\energy_3-\energy_4) \\
    &\quad\times |\mathcal{M}_{12\to 34}|^2\,F(f) \\
    &\quad\times \int d\Omega_2\,d\Omega_3\,d\Omega_4\,
    \delta^{(3)}(\bp_1 \!+\! \bp_2 \!-\! \bp_3 \!-\! \bp_4)\;.
  \end{split}
\end{equation}
The angular integration is the same as Eq.~\eqref{eq:anguler_01} and
the density of states changes as
\begin{equation}
  \rho = \frac{\sqrt{2m\energy}}{2\pi^2}\;.
\end{equation}
After all, we arrive at
\begin{equation}
  \rho_1\frac{\partial f}{\partial t} = g^2\int d\energy_2\,
  d\energy_3\,d\energy_4\,\delta(\energy_1+\energy_2-\energy_3-\energy_4)\,
  F[f]\,\energy_{\min}^{1/2}\;
\end{equation}
For more general discussions, readers can consult
Refs.~\cite{PhysRevA.56.575,PhysRevA.53.381} and, for the mathematical
analysis of the collision integral in the Boltzmann equation, see a
review~\cite{escobedo2003}.

\section{More details on the numerical procedure}
\label{app:chain-rule}

To evaluate the right-hand side of
Eq.~\eqref{eq:chain-rule-eq-for-parameter-flow} we utilize the
following form of the integrations;
\begin{align}
  \begin{split}
    & I[\bar{F}(\energy_1, \energy_2, \energy_2, \energy_3)]
    \equiv \int_0^1 \frac{ds}{s^2} \int_0^1 \frac{dt}{t^2} \\
    &\quad  \times \bigl[\BEP{1}^{-\alpha-3}
      \bar{F}(\BEP{2}/\BEP{1},\BEP{3}/\BEP{1},\BEP{4}/\BEP{1}; \energy_1) \\
    &\quad + \BEP{2}^{-\alpha-3}
      \bar{F}(\BEP{1}/\BEP{2},\BEP{3}/\BEP{2},\BEP{4}/\BEP{2}; \energy_1) \\
    &\quad + \BEP{3}^{-\alpha-3}
      \bar{F}(\BEP{4}/\BEP{3},\BEP{1}/\BEP{3},\BEP{2}/\BEP{3}; \energy_1) \\
    &\quad + \BEP{4}^{-\alpha-3}
      \bar{F}(\BEP{3}/\BEP{4},\BEP{2}/\BEP{4},\BEP{1}/\BEP{4}; \energy_1)
      \bigr]\;,
  \end{split}
\end{align}
where the integrand is a function of $s$ and $t$ with
$\BEP{1}=1$, $\BEP{2}=1/s + 1/t - 1$, $\BEP{3}=1/s$, and $\BEP{4}=1/t$.
For the numerical integration we employed the Gauss-Legendre
quadrature.  To check the convergence, we compared $128^{\rm th}$ and
$256^{\rm th}$ order quadratures.  Using this integration we can write
the right-hand side of Eq.~\eqref{eq:chain-rule-eq-for-parameter-flow}
as
\begin{align}
  \dot{f}(\energy_1)
    &= \energy_1^{2 + \alpha - \delta} I[\bar F]\;,
  \label{eq:app2-collision-1}\\
  \dot{f}'(\energy_1)
  &= \energy_1^{2 + \alpha - \delta} \bigl[
    (2+\alpha-\delta)I[\bar{F}] + I[\bar{F}_x] \bigr]\;,
  \label{eq:app2-collision-2}
\end{align}
where $\bar{F}_x \equiv \partial_x \bar{F}(\BEP{2},\BEP{3},\BEP{4};\energy_1)$
with $x = \ln\energy_1$.  In the $f^3$-regime, the explicit forms of
$\bar{F}$ and $\bar{F}_x$ are
\begin{align}
  \bar{F} = \bar{F}^{(3)} & \equiv f_3 f_4(f_1 + f_2)
      - f_1 f_2(f_3 + f_4)\;, \\
  \begin{split}
  \bar{F}_x = \bar{F}^{(3)}_x & \equiv \xi f_3 f_4 \bigl[
    (E + \BEP{3} n_3 + \BEP{4} n_4)(f_1 + f_2) \bigr] \\
  &\; - \xi f_1 f_2 \bigl[
    (E + \BEP{1} n_1 + \BEP{2} n_2) (f_3 + f_4) \bigr] \\
  &\; + \xi f_3 f_4 \bigl[
    \BEP{1} f_1 (n_1 + 1) + \BEP{2} f_2 (n_2 + 1) \bigr] \\
  &\; - \xi f_1 f_2 \bigl[
    \BEP{3} f_3 (n_3 + 1) + \BEP{4} f_4 (n_4 + 1) \bigr]\;,
  \end{split}
\end{align}
where we defined $\xi \equiv -\beta\energy_1\gamma$,
$E \equiv \BEP{1} + \BEP{2} = \BEP{3} + \BEP{4}$, and
$n_i \equiv 1/(e^{\beta\energy_1\BEP{i}}-1)$.  Likewise, in the
$f^2$-regime, the explicit forms read,
\begin{align}
  \bar{F} = \bar{F}^{(2)} & \equiv f_3 f_4 - f_1 f_2\;, \\
  \begin{split}
  \bar{F}_x = \bar{F}^{(2)}_x &\equiv g f_3 f_4 (E + \BEP{3}n_3 + \BEP{4}n_4) \\
  &\quad -g f_1 f_2 (E + \BEP{1}n_1 + \BEP{2}n_2)\;.
  \end{split}
\end{align}
In the full quantum case, we can just combine expressions in the-$f^3$
regime and the $f^2$-regime to have,
\begin{align}
  \bar{F} &= \bar{F}^{(3)} + \bar{F}^{(2)}\;, \\
  \bar{F}_x &= \bar{F}^{(3)}_x + \bar{F}^{(2)}_x\;.
\end{align}

We can obtain the matrix elements in the left-hand side of
Eq.~\eqref{eq:chain-rule-eq-for-parameter-flow} by taking the
differentiation of the fitting function explicitly as
\begin{align}
  \frac{\partial f(\energy_1)}{\energy_1\partial\beta}
  &= -\gamma f_1 (n_1 + 1)\;, \\
  \frac{\partial f'(\energy_1)}{\energy_1\partial\beta}
  &= -\gamma f_1 (n_1 + 1) \bigl\{ 1 - \beta\energy_1
       [(\gamma + 1) n_1 + \gamma] \bigr\}\;, \\
  \frac{\partial f(\energy_1)}{\partial\gamma}
  &= f_1 \ln n_1\;,\\
  \frac{\partial f'(\energy_1)}{\partial\gamma}
  &= -\beta\energy_1 f_1 (n_1 + 1) (\ln f_1 + 1)\;.
\end{align}

Finally, here, we notice that the collision integrals in
Eqs.~\eqref{eq:app2-collision-1} and \eqref{eq:app2-collision-2} are
functions of $\beta\energy_1$ apart from a common factor
$\energy_1^{2 + \alpha - \dos}$.  The matrix elements in the left-hand
side are also functions of $\beta\energy_1$.  The common factor,
$\energy_1^{2 + \alpha - \dos}$, does not affect the direction of the
flow, but only changes its velocity.  Therefore, only the
$\beta\energy_1$ dependence is relevant for our analysis on the
structure of the flow diagrams.

\bibliographystyle{utphys}
\bibliography{NFP}

\end{document}